\documentclass[preprint,11pt]{elsarticle}

 \usepackage{epsfig}

\usepackage{amssymb}
\usepackage{graphics,amssymb,amsmath}
\usepackage{graphicx}
\usepackage{subfigure}
\newproof{pf}{Proof}
\newdefinition{rmk}{Remark}
\newtheorem{thm}{Theorem}
\newdefinition{prop}{Proposition}
\journal{Computational Statistics and Data Analysis }

\begin{document}

\begin{frontmatter}



\title{A new two parameter lifetime distribution: model and properties }

\author{Hojjatollah Zakerzadeh}
\author{Eisa Mahmoudi\corref{cor1}}
\ead{emahmoudi@yazduni.ac.ir}


\address{Department of Statistics, Yazd University,
P.O. Box 89175-741, Yazd, Iran}

\begin{abstract}
In this paper a new lifetime distribution which is obtained by
compounding Lindley and geometric distributions, named
Lindley-geometric (LG) distribution, is introduced. Several
properties of the new distribution such as density, failure rate,
mean lifetime, moments, and order statistics are derived.
Furthermore, estimation by maximum likelihood and inference for
large sample are discussed. The paper is motivated by two
applications to real data sets and we hope that this model be able
to attract wider applicability in survival and reliability.
\end{abstract}


\end{frontmatter}


\section{Introduction}
The Lindley distribution specified by the probability density
function (p.d.f.)
\begin{equation}\label{pdf lindely}
f(x) = \frac{\theta^{2}}{\theta+1} (1 + x) e^{-\theta
x},~~x>0,~~\theta>0,
\end{equation}
was introduced by Lindley \cite{Lindley }. The corresponding
cumulative distribution function (c.d.f.) is given by
\begin{equation}\label{cdf lindely}
F(x) = 1 -(1+\frac{\theta x}{\theta+1})e^{-\theta
x},~~x>0,~~\theta>0.
\end{equation}
The Lindley distribution, in spite of  little attention in the
statistical literature, is important for studying stress-strength
reliability modeling. Besides, some researchers have proposed new
classes of distributions based on modifications of the Lindley
distribution, including also their properties. Sankaran
\cite{Sankaran} introduced the discrete Poisson-Lindley distribution
by combining the Poisson and Lindley distributions. Ghitany et al.
\cite{Ghitany2008} investigated most of the statistical properties
of the Lindley distribution, showing this distribution may provide a
better fitting than the exponential distribution. Mahmoudi and
Zakerzadeh \cite{Mahmoudi2010 } proposed an extended version of the
compound Poisson distribution which was obtained by compounding the
Poisson distribution with the generalized Lindley distribution which
is obtained and analyzed by Zakerzadeh and Dolati \cite{Zakerzadeh}.
Recently a new extension of the Lindley distribution, called
extended Lindley (EL) distribution, which offers a more flexible
model for lifetime data is introduced by Bakouch et al.
\cite{Bakoucha2011}.

Adamidis and Loukas \cite{Adamidis } introduced a two-parameter
lifetime distribution with decreasing failure rate by compounding
exponential and geometric distributions, which was named exponential
geometric (EG) distribution. In the same way, Kus \cite{Kus } and
Tahmasbi and Rezaei \cite{Tahmasbi } introduced the exponential
Poisson (EP) and exponential logarithmic distributions,
respectively. Marshall and Olkin \cite{Marshall} presented a method
for adding a parameter to a family of distributions with application
to the exponential and Weibull families.

Recently, Chahkandi and Ganjali \cite{Chahkandi} introduced a class
of distributions, named exponential power series (EPS)
distributions, by compounding exponential and power series
distributions, where compounding procedure follows the same way that
was previously carried out by Adamidis and Loukas \cite{Adamidis };
this class contains the distributions mentioned before. Extensions
of the EG distribution was given by Adamidis et al.
\cite{Adamidis2005 } and Barreto-Souza et al.
\cite{Barreto-Souza2011}, where the last was obtained by compounding
Weibull and geometric distributions. A three-parameter extension of
the EP distribution was obtained by Barreto-Souza and Cribari-Neto
\cite{Barreto-Souza2009 }.

This new class of distributions has been received considerable
attention over the two last years. Weibull power series (WPS),
complementary exponential geometric (CEG), two-parameter
Poisson-exponential, generalized exponential power series (GEPS),
exponentiated Weibull-Poisson (EWP) and generalized inverse
Weibull-Poisson (GIWP) distributions were introduced and studied by
Morais and Barreto-Souza \cite{Morais }, Louzada-Neto et al.
\cite{Louzada}, Cancho et al. \cite{Cancho }, Mahmoudi and Jafari
\cite{Mahmoudi2011a }, Mahmoudi and Sepahdar \cite{Mahmoudi2011b }
and Mahmoudi and Torki \cite{Mahmoudi2011c }.

In this paper, we introduce a new lifetime distribution by
compounding Lindley and geometric distributions as follows: Consider
the random variable $X$ having the Lindley distribution where
its pdf and cdf are given in (\ref{pdf lindely}) and (\ref{cdf lindely}).\\
Given $N$, let $X_{1},\cdots,X_{N}$ be independent and identically
distributed (iid) random variables from Lindley distribution. Let
the random variable $N$ is distributed according to the geometric
distribution with pdf
\begin{equation*}
P(N=n)=(1-p)p^{n-1},~n=1,2 ,\cdots,~0<p<1.
\end{equation*}
Let $Y=\min(X_{1},\cdots,X_{N})$, then the conditional cdf of
$Y|N=n$ is given  by
\begin{equation}\label{dist y given N}
F_{Y|N}(y|n)=1-\Big[(1+\frac{\theta y}{\theta+1})e^{-\theta
y}\Big]^{n},
\end{equation}
The Lindley-geometric (LG) distribution, denoted by LG$(p,\theta)$,
is defined by the marginal cdf of $Y$, i.e.,
\begin{equation}\label{cdf LG}
F_Y(y)=\frac{1-(1+\frac{\theta y}{\theta+1})e^{-\theta
y}}{1-p(1+\frac{\theta y}{\theta+1})e^{-\theta
y}},~~y>0,~~\theta>0,~~0< p <1.
\end{equation}

The paper is organized as follows. In Section 2, the density
function, survival and hazard rate functions of the LG with some of
their properties are given. Section 3 provides a general expansion
for the quantiles and moments of the LG distribution. Its moment
generating function is derived in this section. Section 4 provides
the moments of order statistics of the LG distribution. Residual
life and reversed residual functions of the LG distribution is
discussed in Section 5. Section 6 is devoted to the Bonferroni and
Lorenz curves of the LG distribution. In Section 7 we explain the
probability weighted moments. Mean deviations from the mean and
median are derived in Section 8. Estimation of the parameters by
maximum likelihood via an EM-algorithm and inference for large
sample are presented in Section 9. Applications to two real data
sets are given in Section 10 and conclusions are provided in Section
11.

\section{Density function, survival and hazard rate functions}

The probability density function of the LG distribution is given by
\begin{equation}\label{pdf LG}
f(y)=\frac{\theta^2}{\theta+1}(1-p)(1+y)e^{-\theta
y}\Big[1-p(1+\frac{\theta y}{\theta+1})e^{-\theta
y}\Big]^{-2},~~y>0,
\end{equation}
where $\theta>0$ and $0<p<1$.\\
Even when $p\leq0$, Equation (\ref{pdf LG}) is a density function.
We can then define the LG distribution by Equation (\ref{pdf LG})
for any $p<1$. Some special sub-models of the LG distribution
(\ref{pdf LG}) are obtained as follows. If $p = 0$, we have the
Lindley distribution. When $p\rightarrow 1^{-}$, the LG distribution
tends to a distribution degenerate in zero. Hence, the parameter $p$
can be interpreted as a concentration parameter. LG density
functions are displayed in Figure 1 for selected values of $\theta$
and $p = -2,~-.05,~ 0,~ .05,~ .09$.
\begin{thm}
The density function of the LG distribution is (i) decreasing for
all values $p$ and $\theta$ for which $p>
\frac{1-\theta^2}{1+\theta^2}$, (ii) unimodal for all values $p$ and
$\theta$ for which $p\leq \frac{1-\theta^2}{1+\theta^2}.$
\end{thm}
\begin{pf}
See Appendix.
\end{pf}

The survival function and hazard rate function of the LG
distribution, are given respectively by
\begin{equation}\label{Survival}
S(y)=\frac{(1-p)(1+\frac{\theta y}{\theta+1})e^{-\theta
y}}{1-p(1+\frac{\theta y}{\theta+1})e^{-\theta y}},
\end{equation}
and
\begin{equation}\label{hazard}
h(y)=\frac{\theta^{2}(y+1)}{\theta
y+\theta+1}\Big[1-p(1+\frac{\theta y}{\theta+1})e^{-\theta
y}\Big]^{-1}.
\end{equation}
Figure 2 provides the plots of the hazard rate function of the LG
distribution for different values $p = -2,~-.05,~ 0,~ .05,~ .09$ and
$\theta=.03,~.08,~1,~3$. We have the following results regarding the
shapes of the hazard rate function of the LG distribution. The proof
is provided in the Appendix.
\begin{thm}
The hazard function of the LG distribution in (\ref{hazard}) is (i)
bathtub-shaped if $p>\frac{1}{1+\theta^2},$ (ii) firstly increasing
then bathtub-shaped if $p\leq\frac{1}{1+\theta^2}.$
\end{thm}

\begin{prop}
The hazard rate function of the LG distribution in (\ref{hazard})
tends to $\frac{\theta^{2}}{(\theta+1)(1-p)}$ and $\theta$ where
$y\rightarrow 0$ and $y\rightarrow \infty$, respectively.
\end{prop}

Using the series expansion
\begin{equation}\label{series}
(1-z)^{-k}=\sum_{j=0}^{\infty}\frac{\Gamma(k+j)}{\Gamma(k)j!}z^j,
\end{equation}
where $|z|< 1$ and $k > 0$, the density function (\ref{pdf LG}) can
be demonstrated by
\begin{equation}\label{new LG}
f_{LG}(y;p,\theta)=\frac{\theta^2}{\theta+1}(1-p)(1+y)e^{-\theta
y}\sum_{j=0}^{\infty}(j+1)p^{j}(1+\frac{\theta
y}{\theta+1})^{j}e^{-j\theta y}.
\end{equation}
Various mathematical properties of the LG distribution can be
obtained from (\ref{new LG}) and the corresponding properties of the
Lindely distribution.\\
In the following theorem, we give the stochastically ordering
property of the random variable $Y$ with LG distribution.
\begin{thm}
Consider the two random variables $Y_1$ and $Y_2$ with
$LG(p_{1},\theta)$ and $LG(p_{2},\theta)$ distributions,
respectively.
\begin{description}
    \item[(i)] If $p_{1}\leq p_{2},$ then $S_{Y_{1}}(t)\leq
    S_{Y_{2}}(t)$ ($Y_{1}\leq
    Y_{2}$) and $h_{Y_{1}}(t)\leq
    h_{Y_{2}}(t)$ ($Y_{1}\leq
    Y_{2}$).
    \item[(ii)] If $p_{1}>p_{2},$ then
    $\frac{f_{Y_{1}}(t)}{f_{Y_{2}}(t)}$ is decreasing in $t$, i.e., ($Y_{1}\leq
    Y_{2}$).
\end{description}
\end{thm}

\section{ Quantiles and moments of the LG distribution }

Applying the equation $F(x_{\xi})=\xi$, the $\xi$th quantile of the
LG distribution is the solution of equation
$$\frac{1-\xi}{1-p\xi}=(1+\frac{\theta x_{\xi}}{\theta+1})e^{-\theta
x_{\xi}},$$ which is used for data generation from the LG
distribution.\\
Suppose that $Y\sim LG(p,\theta)$, using the equation (\ref{new LG})
and applying the binomial expression for $(1+\frac{\theta
y}{\theta+1})^{j}$, the $r$th moment of $Y$ is given by
\begin{equation}\label{mean r LG}
E(Y^{r})=\frac{\theta^{2}(1-p)}{\theta+1}\sum_{j=0}^{\infty}\sum_{i=0}^{j}{j
\choose
i}(j+1)p^{j}(\frac{\theta}{\theta+1})^{i}\frac{\Gamma(r+i+1)}{(\theta(j+1))^{r+i+1}}(1+\frac{r+i+1}{\theta(j+1)}).
\end{equation}
Using Eq. (\ref{mean r LG}), the moment generating function of the
LG distribution is given by
\begin{equation*}
\begin{array}[b]{l}\label{mgf EWP}
M_{Y}(t)=\sum^{\infty}_{k=0}\frac{t^k}{k!}\Big[\frac{\theta^{2}(1-p)}{\theta+1}\sum_{j=0}^{\infty}\sum_{i=0}^{j}{j
\choose
i}(j+1)p^{j}(\frac{\theta}{\theta+1})^{i}\frac{\Gamma(k+i+1)}{(\theta(j+1))^{k+i+1}}(1+\frac{k+i+1}{\theta(j+1)})\Big]\medskip\\
~~~~~~~~=\frac{\theta^{2}(1-p)}{\theta+1}\sum^{\infty}_{k=0}\sum_{j=0}^{\infty}\sum_{i=0}^{j}\frac{t^k}{k!}{j\choose
i}(j+1)p^{j}(\frac{\theta}{\theta+1})^{i}\frac{\Gamma(k+i+1)}{(\theta(j+1))^{k+i+1}}(1+\frac{k+i+1}{\theta(j+1)}).
\end{array}
\end{equation*}

\begin{prop}
The mean of the LG distribution ig given by
\begin{equation}\label{mean LG}
E(Y)=\frac{\theta^{2}(1-p)}{\theta+1}\sum_{j=0}^{\infty}\sum_{i=0}^{j}\frac{(j+1)!}{(j-i)!}p^{j}
(\frac{\theta}{\theta+1})^{i}\frac{i+1}{(\theta(j+1))^{i+2}}(1+\frac{i+2}{\theta(j+1)}).
\end{equation}
\end{prop}

\section{Order statistics and their moments}
Order statistics make their appearance in many areas of statistical
theory and practice. Order statistics are among the most fundamental
tools in non-parametric statistics and inference and play an
important role in quality control testing and reliability, where a
practitioner needs to predict the failure of future items based on
the times of a few early failures.\\ Let $Y_{1}, \cdots, Y_{n}$ be a
random sample taken from the LG distribution and $Y_{1:n}, \cdots,
Y_{n:n}$ denote the corresponding order statistics. Then, the pdf
$f_{r:n}(y)$ of the \textit{r}th order statistics $Y_{r:n}$ is given
by
\begin{equation*}
\begin{array}[b]{l}
f_{r:n}(y)=\frac{1}{Be(r,n-r+1)}\frac{\theta^2}{\theta+1}(1-p)(1+y)e^{-\theta
y}\sum_{j=0}^{\infty}\sum_{i=0}^{r}\sum_{l=0}^{\infty}(-1)^{i}{r\choose
i}\frac{\Gamma(n-r+i+l)}{l!}(j+1)p^{j+l}\medskip\\
~~~~~~~~~\times (1-p)^{n-r+i}\Big(1+\frac{\theta
y}{\theta+1}\Big)^{n-r+i+j+l}e^{-(n-r+i+j+l)\theta y},
\end{array}
\end{equation*}
where $Be(a, b) =\int_{0}^{1}w^{a-1}(1-w)^{b-1}dw $ is the beta
function. After some calculations and using the binomial expression
for $\Big(1+\frac{\theta y}{\theta+1}\Big)^{n-r+i+j+l}$, we have
\begin{equation*}
\begin{array}[b]{l}
f_{r:n}(y)=\frac{1}{Be(r,n-r+1)}\frac{\theta^2}{\theta+1}(1-p)(1+y)e^{-\theta
y}\sum_{j=0}^{\infty}\sum_{i=0}^{r}\sum_{l=0}^{\infty}\sum_{m=0}^{n-r+i+j+l}(-1)^{i}{n-r+i+j+l\choose
m}\medskip\\
~~~~~~~~~\times {r\choose
i}\frac{\Gamma(n-r+i+l)}{l!}(j+1)p^{j+l}(1-p)^{n-r+i}\Big(\frac{\theta
y}{\theta+1}\Big)^{m}e^{-(n-r+i+j+l)\theta y}.
\end{array}
\end{equation*}
The \textit{k}th moment of the \textit{r}th order statistic
$Y_{r:n}$ can be obtained from the known result,
\begin{equation*}
\begin{array}[b]{l}
E[Y^{k}_{r:n}]=\frac{1}{Be(r,n-r+1)}\frac{\theta^2}{\theta+1}(1-p)\sum_{j=0}^{\infty}\sum_{i=0}^{r}\sum_{l=0}^{\infty}\sum_{m=0}^{n-r+i+j+l}(-1)^{i}{n-r+i+j+l\choose
m}{r\choose
i}\frac{\Gamma(n-r+i+l)}{l!}\medskip\\
~~~~~~~~~\times (j+1)p^{j+l}(1-p)^{n-r+i}\Big(\frac{\theta
}{\theta+1}\Big)^{m}\left[
\frac{\Gamma(k+m+2)}{(\theta(n-r+i+j+l+1))^{k+m+2}}+\frac{\Gamma(k+m+1)}{(\theta(n-r+i+j+l+1))^{k+m+1}}\right].
\end{array}
\end{equation*}

\section{ Residual life and reversed failure rate function of the LG distribution}

Given that a component survives up to time $t>0$, the residual life
is the period beyond $t$ until the time of failure and defined by
the conditional random variable $Y - t|Y > t$. In reliability, it is
well known that the mean residual life function and ratio of two
consecutive moments of residual life determine the distribution
uniquely (Gupta and Gupta, \cite{Gupta1983}). Therefore, we obtain
the $r$th-order moment of the residual life via the general formula
\begin{equation*}
\mu_{r}(t)=E\left[(Y-t)^r|Y>t\right]=\frac{1}{S(t)}\int_{t}^{\infty}(y-t)^{r}f(y)dy,
\end{equation*}
where $S(t)=1-F(t)$, is the survival function. \\Applying the
binomial expansion to $(y-t)^r$ into the above formula gives
\begin{equation}
\begin{array}[b]{ll}
\mu_{r}(t)=&\frac{(1-p)\theta^{2}}{(\theta+1)S(t)}\sum_{i=0}^{r}\sum_{j=0}^{\infty}\sum_{k=0}^{j}(-1)^{i}{j\choose
k}(j+1)t^{i}p^{j}\Big(\frac{\theta}{\theta+1}\Big)^{k}\Big[\frac{1}{(\theta(j+1))^{r+k-i+2}}\medskip\\
&\times\Big(\Gamma(r+k-i+2;\theta
t(j+1))+\theta(j+1)\Gamma(r+k-i+1;\theta
t(j+1))\Big)\Big],~~~~~r\geq1,
\end{array}
\end{equation}
where $\Gamma(s;t)= \int_{t}^{\infty}x^{s-1}e^{-x}dx,$ shows the
upper incomplete gamma function.\\ The mean residual life (MRL) of
the LG distribution is given by
\begin{equation}
\begin{array}[b]{ll}
\mu(t)=&\frac{(1-p)\theta^{2}}{(\theta+1)S(t)}\sum_{j=0}^{\infty}\sum_{k=0}^{j}{j\choose
k}(j+1)p^{j}\Big(\frac{\theta}{\theta+1}\Big)^{k}\Big[\frac{1}{(\theta(j+1))^{k+3}}\medskip\\
&\times\Big(\Gamma(k+3;\theta t(j+1))+\theta(j+1)\Gamma(k+2;\theta
t(j+1))\Big)\Big]-t.
\end{array}
\end{equation}
In particular, we obtain
\begin{equation}
\mu(0)=E(Y)=\frac{\theta^{2}(1-p)}{\theta+1}\sum_{j=0}^{\infty}\sum_{i=0}^{j}\frac{(j+1)!}{(j-i)!}p^{j}
(\frac{\theta}{\theta+1})^{i}\frac{i+1}{(\theta(j+1))^{i+2}}(1+\frac{i+2}{\theta(j+1)}).
\end{equation}
Also, if $p=0$, then $$\mu(t)=\frac{2+\theta+\theta
t}{\theta(1+\theta+\theta t)},$$ which is the MRL function of the
original Lindley distribution. The variance of the residual life of
the LG distribution can be obtained easily by using $\mu_{2}(t)$ and
$\mu(t)$.

The reversed residual life can be defined as the conditional random
variable $t-Y|Y\leq t$ which denotes the time elapsed from the
failure of a component given that its life is less than or equal to
$t$. This random variable may also be called the inactivity time (or
time since failure); for more details one can see (Kundu and Nanda,
\cite{Kundu2010}; Nanda et al., \cite{Nanda}). Also, in reliability,
the mean reversed residual life and ratio of two consecutive moments
of reversed residual life characterize the distribution uniquely.
Using (\ref{cdf LG}) and (\ref{pdf LG}), the reversed failure (or
reversed hazard) rate function is given by
\begin{equation}
r(y)=\frac{f(y)}{F(y)}=\frac{\frac{\theta^2}{\theta+1}(1-p)(1+y)e^{-\theta
y}}{\Big[1-p(1+\frac{\theta y}{\theta+1})e^{-\theta
y}\Big]\Big[1-(1+\frac{\theta y}{\theta+1})e^{-\theta
y}\Big]},~~~y>0.
\end{equation}
It is noticed that $h(0) =\infty $ and $h(0)$ is discontinuous in
the parameters of the LG distribution. The \textit{r}th-order moment
of the reversed residual life can be obtained by the well known
formula
\begin{equation*}
m_{r}(t)=E\left[(t-Y)^r|Y\leq
t\right]=\frac{1}{F(t)}\int_{0}^{t}(t-y)^{r}f(y)dy,
\end{equation*}
hence,
\begin{equation}
\begin{array}[b]{ll}
m_{r}(t)=&\frac{(1-p)\theta^{2}}{(\theta+1)F(t)}\sum_{i=0}^{r}\sum_{j=0}^{\infty}\sum_{k=0}^{j}(-1)^{r+i}{j\choose
k}(j+1)t^{i}p^{j}\Big(\frac{\theta}{\theta+1}\Big)^{k}\Big[\frac{1}{(\theta(j+1))^{r+k-i+2}}\medskip\\
&\times\Big(\gamma(r+k-i+2;\theta
t(j+1))+\theta(j+1)\gamma(r+k-i+1;\theta
t(j+1))\Big)\Big],~~~~~r\geq1,
\end{array}
\end{equation}
where $\gamma(s;t)= \int_{0}^{t}x^{s-1}e^{-x}dx,$ shows the lower
incomplete gamma function. Thus, the mean of the reversed residual
life of the LG distribution is given by

\begin{equation}
\begin{array}[b]{ll}
m(t)=&t-\frac{(1-p)\theta^{2}}{(\theta+1)S(t)}\sum_{j=0}^{\infty}\sum_{k=0}^{j}{j\choose
k}(j+1)p^{j}\Big(\frac{\theta}{\theta+1}\Big)^{k}\Big[\frac{1}{(\theta(j+1))^{k+3}}\medskip\\
&\times\Big(\gamma(k+3;\theta t(j+1))+\theta(j+1)\gamma(k+2;\theta
t(j+1))\Big)\Big].
\end{array}
\end{equation}
Using $m(t)$ and $m_{2}(t)$ one can obtain the variance and the
coefficient of variation of the reversed residual life of the LG
distribution.

\section{Bonferroni and Lorenz curves of the LG distribution}

The Bonferroni and Lorenz curves and Gini index have many
applications not only in economics to study income and poverty, but
also in other fields like reliability, medicine and insurance. The
Bonferroni curve $B_{F} [F (y)]$ is given by $$B_{F} [F (y)] =
\frac{1}{\mu F(y)}\int_{0}^{y} uf (u)du.$$ Using this fact that
\begin{equation}
\begin{array}[b]{ll}
I(y)=\int_{0}^{y} uf
(u)du=&\frac{(1-p)\theta^{2}}{(\theta+1)}\sum_{j=0}^{\infty}\sum_{k=0}^{j}{j\choose
k}(j+1)p^{j}\Big(\frac{\theta}{\theta+1}\Big)^{k}\Big[\frac{1}{(\theta(j+1))^{k+3}}\medskip\\
&\times\Big(\gamma(k+3;\theta y(j+1))+\theta(j+1)\gamma(k+2;\theta
y(j+1))\Big)\Big],
\end{array}
\end{equation}
the Bonferroni curve of the distribution function $F$ of LG
distribution is given by
\begin{equation}
\begin{array}[b]{ll}
B_{F} [F (y)]=&\frac{(1-p)\theta^{2}[1-p(1+\frac{\theta
y}{\theta+1})e^{-\theta y}]}{\mu(\theta+1)[1-(1+\frac{\theta
y}{\theta+1})e^{-\theta
y}]}\sum_{j=0}^{\infty}\sum_{k=0}^{j}{j\choose
k}(j+1)p^{j}\Big(\frac{\theta}{\theta+1}\Big)^{k}\Big[\frac{1}{(\theta(j+1))^{k+3}}\medskip\\
&\times\Big(\gamma(k+3;\theta y(j+1))+\theta(j+1)\gamma(k+2;\theta
y(j+1))\Big)\Big],
\end{array}
\end{equation}
where $\mu$ (the mean of LG distribution) is given in (\ref{mean
LG}).\\Also, the Lorenz curve of $F$ that follows the LG
distribution can be obtained via the expression $L_{F} [F (y)] =
B_{F} [F (y)]F (y)$. The scaled total time and cumulative total time
on test transform of a distribution function $F$ (Pundir et al.,
\cite{Pundir} are defined by $$S_{F} [F (t)] = \frac{1}{\mu
}\int_{0}^{t} S (u)du,$$ and $$C_{F}= \int_{0}^{1} S_{F} [F
(t)]f(t)dt,$$ respectively, where $S(.)$ denotes the survival
function. If $F (t)$ denotes the LG distribution function specified
by (\ref{cdf LG}) then we have,
$$S_{F} [F (t)] =\frac{1-p}{\mu}\sum_{j=0}^{\infty}\sum_{k=0}^{j+1}{j+1\choose
k}p^{j}\Big(\frac{\theta}{\theta+1}\Big)^{k}(\theta(j+1))^{-(k+1)}\gamma(k+1;\theta
t(j+1)).$$ The Gini index can be obtained from the relationship $G =
1 - C_F$ .

\section{Probability weighted moments}
The probability weighted moments (PWMs) method can generally be used
for estimating parameters of a distribution whose inverse form
cannot be expressed explicitly. We calculate the PWMs of the LG
distribution since they can be used to obtain the moments of the LG
distribution. The PWMs of a random variable $Y$ are formally defined
by
\begin{equation}\label{PWMs1}
\tau _{s,r}=E[Y^{s}F(Y)^{r}]=\int\limits_{0}^{\infty
}y^{s}F(y)^{r}f(y)dy,
\end{equation}
where $r$ and $s$ are positive integers and $F(.)$ and $f(.)$ are
the cdf and pdf of the random variable $Y$. The PWMs of the LG
distribution are given in the following proposition.

\begin{prop}
The PWMs of the LG distribution with cdf (\ref{cdf LG}) and pdf
(\ref{pdf LG}), are given by
\begin{equation}\label{PWMs2}
\begin{array}[b]{ll}
\tau_{s,r}&=\frac{\theta^{m+2}(1-p)}{(\theta+1)^{m+1}}\sum_{j=0}^{\infty}\sum_{k=0}^{\infty}\sum_{l=0}^{r}\sum_{m=0}^{k+j+l}(-1)^{l}{r\choose
l}{k+j+l\choose m}{r+k-1\choose k}(j+1)p^{j+k}\medskip \\
&~~~\times\left[\frac{\Gamma(m+s+1)}{(\theta(k+j+l))^{m+s+1}}-\frac{\Gamma(m+s+2)}{(\theta(k+j+l))^{m+s+2}}\right].
\end{array}
\end{equation}
\end{prop}

\begin{prop}
The $s$th moment of the LG distribution can be obtained putting
$r=0$ in Eq. (\ref{PWMs2}). Also, the mean and variance of the LG
distribution can be obtained.
\end{prop}

\section{ Mean deviations}
The amount of scatter in a population can be measured by the
totality of deviations from the mean and median. For a random
variable $X$ with pdf $f(.)$, cdf $F(.)$, mean $\mu= E(X)$ and $ M =
Median(X)$, the mean deviation about the mean and the mean deviation
about the median, are defined respectively by
$$\delta_{1}(X)=\int_{0}^{\infty}|x-\mu|f(x)dx=2\mu F(\mu)-2I(\mu),$$
and
$$\delta_{2}(X)=\int_{0}^{\infty}|x-M|f(x)dx=\mu -
2I(M),$$ where $I(b)=\int^{b}_{0}xf(x)dx$.

For the LG distribution we have
\begin{equation}\label{L(b)}
\begin{array}[b]{ll}
I(b)=&\frac{(1-p)\theta^{2}}{(\theta+1)}\sum_{j=0}^{\infty}\sum_{k=0}^{j}{j\choose
k}(j+1)p^{j}\Big(\frac{\theta}{\theta+1}\Big)^{k}\Big[\frac{1}{(\theta(j+1))^{k+3}}\medskip\\
&\times\Big(\gamma(k+3;\theta b(j+1))+\theta(j+1)\gamma(k+2;\theta
b(j+1))\Big)\Big].
\end{array}
\end{equation}

\begin{thm}
The Mean deviations of the LG distribution are given by
\begin{equation*}
\delta_{1}=2\mu \frac{1-(1+\frac{\theta \mu}{\theta+1})e^{-\theta
\mu}}{1-p(1+\frac{\theta \mu}{\theta+1})e^{-\theta \mu}}-2I(\mu),
\end{equation*}
and
\begin{equation*}
\delta_{2}=\mu-2I(M),
\end{equation*}
respectively, where $\mu$ is the mean of LG in Eq. (\ref{mean LG}),
$I(\mu)$ and $I(M)$ are obtained by substituting $\mu$ and $M$ in
Eq. (\ref{L(b)}).
\end{thm}

\section{ Estimation and inference}
The estimation of the parameters of the LG distribution using the
maximum likelihood estimation is studied in this section. Let
$Y_{1},Y_{2},\cdots,Y_{n}$ be a random sample with observed values
$y_{1},y_{2},\cdots,y_{n}$ from LG distribution with parameters $p$
and $\theta$. The total log-likelihood function is given by
\begin{equation*}
\begin{array}[b]{ll}
l_{n}(y;p,\theta)&=2n\log(\theta)-n\log(1+\theta)+n\log(1-p)+\sum_{i=1}^{n}\log(1+y_{i})-\theta\sum_{i=1}^{n}y_{i}\medskip\\
&~~~- 2\sum_{i=1}^{n}\log\Big(1-p(1+\frac{\theta
y_{i}}{\theta+1})e^{-\theta y_{i}}\Big).
\end{array}
\end{equation*}
The associated score function is given by $U_{n}=(\partial
l_{n}/\partial p,\partial l_{n}/\partial \theta)^{T}$, where

\begin{equation*}
\begin{array}{ll}
\frac{\partial l_{n}}{\partial
p}&=\frac{-n}{1-p}+2\sum_{i=1}^{n}\frac{(1+\frac{\theta
y_{i}}{\theta+1})e^{-\theta y_{i}}}{1-p(1+\frac{\theta
y_{i}}{\theta+1})e^{-\theta y_{i}}},\medskip \\
\frac{\partial l_{n}}{\partial
\theta}&=\frac{2n}{\theta}-\frac{n}{1+\theta}-\sum_{i=1}^{n}y_{i}-2p\sum_{i=1}^{n}\frac{y_{i}e^{-\theta
y_{i}}\Big[(1+\frac{\theta
y_{i}}{\theta+1})-\frac{1}{(\theta+1)^{2}}\Big]}{\Big[1-p(1+\frac{\theta
y_{i}}{\theta+1})e^{-\theta y_{i}}\Big]^{2}}.
\end{array}
\end{equation*}
The maximum likelihood estimation (MLE) of $p $ and $\theta$ is
obtained by solving the nonlinear system $U_n=\textbf{0}$. The
solution of this nonlinear system of equation has not a closed form.
For interval estimation and hypothesis tests on the model
parameters, we require the information matrix.

\subsection{Asymptotic variances and covariances of the MLEs}
Applying the usual large sample approximation, MLE of $\Theta$ i.e.
$\widehat{\Theta }=(\hat{p},\hat{\theta})$, can be treated as being
approximately bivariate normal with mean $\Theta$ and
variance-covariance matrix, which is the inverse of the expected
information matrix $J(\Theta)=E[I(\Theta)],$ i.e., $N_2(\Theta
,{J(\Theta )}^{-1})$, where $I(\Theta;y_{obs})$ is the observed
information matrix with elements $I_{ij}=-\partial^{2}l/\partial
\theta_{i}\theta_{j}$ with $i,~j=~1,2$ and the expectation is to be
taken with respect to the distribution of $Y$. Differentiating
$\partial l/\partial p$ and $\partial l/\partial \theta$, the
elements of the symmetric, second-order observed information matrix
are found to be
\begin{equation*}
\begin{array}{ll}
I_{11}&=\frac{n}{(1-p)^2}-2\sum_{i=1}^{n}\frac{\big[(1+\frac{\theta
y_{i}}{\theta+1})e^{-\theta y_{i}}\big]^2}{\big[1-p(1+\frac{\theta
y_{i}}{\theta+1})e^{-\theta y_{i}}\big]^2},\medskip\\
I_{12}&=2\sum_{i=1}^{n}\frac{y_{i}e^{-\theta
y_{i}}\big[(1+\frac{\theta
y_{i}}{\theta+1})-\frac{1}{(\theta+1)^{2}}\big]}{\big[1-p(1+\frac{\theta
y_{i}}{\theta+1})e^{-\theta
y_{i}}\big]^{2}}+4p\sum_{i=1}^{n}\frac{\big[y_{i}e^{-\theta
y_{i}}\big((1+\frac{\theta
y_{i}}{\theta+1})-\frac{1}{(\theta+1)^{2}}\big)\big]\big[(1+\frac{\theta
y_{i}}{\theta+1})e^{-\theta y_{i}}\big]}{\big[1-p(1+\frac{\theta
y_{i}}{\theta+1})e^{-\theta y_{i}}\big]^{3}},\medskip\\
I_{22}&=\frac{2 n}{\theta ^2}-\frac{n}{(1+\theta )^2}+2 p \sum
_{i=1}^n \Big[\frac{y_i e^{-\theta  y_i} \left(\frac{2}{(1+\theta
)^3}-\frac{\theta  y_i}{(1+\theta )^2}+\frac{y_i}{1+\theta
}\right)}{\left(1-p e^{-\theta  y_i} \left(1+\frac{\theta
y_i}{1+\theta }\right)\right){}^2}-\frac{y_i^2e^{-\theta  y_i}
\left(1-\frac{1}{(1+\theta )^2}+\frac{\theta  y_i}{1+\theta
}\right)}{\left(1-p e^{-\theta  y_i} \left(1+\frac{\theta
y_i}{1+\theta }\right)\right){}^2}\medskip\\
&~~-\frac{2y_i  e^{-\theta  y_i} \left(1-\frac{1}{(1+\theta
)^2}+\frac{\theta  y_i}{1+\theta }\right) \left(p e^{-\theta  y_i}
\left(\frac{\theta  y_i}{(1+\theta )^2}-\frac{y_i}{1+\theta
}\right)+p y_i e^{-\theta  y_i} \left(1+\frac{\theta y_i}{1+\theta
}\right)\right)}{\left(1-p e^{-\theta  y_i} \left(1+\frac{\theta
y_i}{1+\theta }\right)\right){}^3}\Big].
\end{array}
\end{equation*}
The elements of the expected information matrix, $J(\Theta),$ are
calculated by taking the expectations of $I_{ij}$, $i,~j=~1,2$, with
respect to the distribution of $Y$. When the expectations of
$I_{ij}$, $i,~j=~1,2$ is obtained, we would have the matrix
$J(\Theta),$ the inverse of $J(\Theta),$ evaluated at
$\widehat{\Theta }$ provides the asymptotic variance-covariance
matrix of MLEs. Alternative estimates can be obtained from the
inverse of the observed information matrix since it is a consistent
estimator of $J^{-1}(\Theta)$.\\
The estimated asymptotic multivariate normal $N_2(\Theta
,{I(\widehat{\Theta })}^{-1})$ distribution of $\widehat{\Theta }$
can be used to construct approximate confidence intervals for the
parameters and for the hazard rate and survival functions. An
$100(1-\gamma )$ asymptotic confidence interval for parameters $p$
and $\theta$ is given by
\[{ACI}_p=(\hat{p
}-Z_{\frac{\gamma }{2}}\sqrt{{\hat{I}}_{pp}},\hat{p
}+Z_{\frac{\gamma }{2}}\sqrt{{\hat{I}}_{pp}}),\] and
\[{ACI}_\theta=(\hat{\theta
}-Z_{\frac{\gamma }{2}}\sqrt{{\hat{I}}_{\theta \theta}},\hat{\theta
}+Z_{\frac{\gamma }{2}}\sqrt{{\hat{I}}_{\theta \theta}}),\] where
${\hat{I}}_{pp}$ and ${\hat{I}}_{\theta \theta}$ are the diagonal
element of ${I(\widehat{\Theta })}^{-1}$ and $Z_{\frac{\gamma }{2}}$
is the quantile $1-\gamma /2$ of the standard normal distribution.

We can compute the maximized unrestricted and restricted
log-likelihoods to construct likelihood ratio (LR) statistics for
testing some LG sub-models. For example, we can use LR statistics to
check whether the fitted LG distribution for a given data set is
statistically "superior" to the fitted Lindely distribution. In any
case, hypothesis tests of the type $H_{0}:~\Theta = \Theta_0$ versus
$H_{1}:~\Theta = \Theta_0$ can be performed using LR statistics. In
this case, the LR statistic for testing $H_{0}$ versus $H_{1}$ is $w
= 2\{l(\widehat{\Theta})- l(\widehat{\Theta}_{0})\}$, where
$\widehat{\Theta}$ and $\widehat{\Theta}_{0}$ are the MLEs under
$H_{1}$ and $H_{0}$. The statistic $w$ is asymptotically (as $n
\rightarrow \infty$) distributed as $\chi^{2}_{k}$, where $k$ is the
dimension of the subset $\Theta$ of interest.

\subsection{An EM algorithm }
Let the complete-data be $Y_{1},\cdots,Y_{n}$ with observed values
$y_{1},\cdots,y_{n}$ and the hypothetical random variable
$Z_{1},\cdots,Z_{n}$. The joint probability density function is such
that the marginal density of $Y_{1},\cdots,Y_{n}$ is the likelihood
of interest. Then, we define a hypothetical complete-data
distribution for each $(Y_{i},Z_{i})~~i=1,\cdots,n$ with a joint
probability density function in the form
\begin{equation}
g(y,z;\Theta)=(1-p)\frac{z\theta^2}{\theta+1}(1+y)e^{-\theta
y}\big[p(1+\frac{\theta y}{\theta+1})e^{-\theta y}\big]^{z-1},
\end{equation}
where $\Theta=(p,\theta)$, $y>0$ and $z\in
\mathbb{N}$.\\
Under the formulation, the E-step of an EM cycle requires the
expectation of $(Z|Y;\Theta^{(r)})$ where
$\Theta^{(r)}=(\alpha^{(r)},\beta^{(r)},\gamma^{(r)},\theta^{(r)})$
is the current estimate (in the $r$th iteration) of $\Theta$.\\
The pdf of $Z$ given $Y$, say $g(z|y)$ is given by
\begin{equation*}
g(z|y)=z\big[p(1+\frac{\theta y}{\theta+1})e^{-\theta
y}\big]^{z-1}\big(1-p(1+\frac{\theta y}{\theta+1})e^{-\theta
y}\big)^2.
\end{equation*}
Thus, its expected value is given by
\begin{equation*}
E[Z|Y=y]=\frac{\big(1+p(1+\frac{\theta y}{\theta+1})e^{-\theta
y}\big)}{\big(1-p(1+\frac{\theta y}{\theta+1})e^{-\theta y}\big)}.
\end{equation*}

The EM cycle is completed  with the M-step by using the maximum
likelihood estimation over $\Theta$, with the missing $Z$'s
replaced by  their conditional expectations given above.\\
The log-likelihood for the complete-data is

\begin{equation*}
\begin{array}{ll}
l^{*}_{n}(\textbf{y},\textbf{z};\Theta)&\propto \sum^{n}_{i=1} \log
z_{i} +
2n\log(\theta)-n\log(1+\theta)+n\log(1-p)-\theta\sum_{i=1}^{n}y_{i}\medskip\\
&~~~+\sum_{i=1}^{n}\log(1+y_{i})+\sum_{i=1}^{n}(z_{i}-1)\log\big(p(1+\frac{\theta
y_{i}}{\theta+1})e^{-\theta y_{i}}\big).
\end{array}
\end{equation*}

The components of the score function $U^{*}_{n}(\Theta)=
(\frac{\partial l^{*}_{n}}{\partial p},\frac{\partial l^{*}_{n}
}{\partial \theta})^{T}$ are given by
\begin{equation*}
\begin{array}{ll}
\frac{\partial
l^{*}_{n}}{\partial p}&=\sum^{n}_{i=1}\frac{z_{i}-1}{p}-\frac{n}{1-p},\medskip\\
\frac{\partial
l^{*}_{n}}{\partial\theta}&=\frac{2n}{\theta}-\frac{n}{\theta+1}-\sum^{n}_{i=1}y_{i}+\sum^{n}_{i=1}(z_{i}-1)\frac{y_{i}e^{-\theta
y_{i}}\big(\frac{1}{(\theta+1)^2}-\frac{\theta
y_{i}}{\theta+1}-1\big)}{(1+\frac{\theta y_{i}}{\theta+1})e^{-\theta
y_{i}}}.
\end{array}
\end{equation*}
From a nonlinear system of equations $U^{*}_{n}(\Theta)=\textbf{0}$,
we obtain the iterative procedure of the EM algorithm as
\begin{equation*}
\begin{array}{l}
\hat p^{(r+1)}=1-\frac{n}{\sum^{n}_{i=1} z_{i}^{(r)}},\medskip\\
\frac{2n}{\hat\theta^{(r+1)}}-\frac{n}{\hat\theta^{(r+1)}+1}-\sum^{n}_{i=1}y_{i}+\sum^{n}_{i=1}
(z_{i}^{(r)}-1)\frac{y_{i}e^{-\hat\theta^{(r+1)}
y_{i}}\big(\frac{1}{(\hat\theta^{(r+1)}+1)^2}-\frac{\hat\theta^{(r+1)}
y_{i}}{\hat\theta^{(r+1)}+1}-1\big)}{\big(1+\frac{\hat\theta^{(r+1)}
y_{i}}{\hat\theta^{(r+1)}+1}\big)e^{-\hat\theta^{(r+1)} y_{i}}},
\end{array}
\end{equation*}
where $\hat\theta^{(r+1)}$ is found numerically. Hence, for
$i=1,\cdots,n$, we have that
\begin{equation*}
z^{(r)}_{i}=\frac{\big(1+\hat p^{(r)}(1+\frac{\hat\theta^{(r)}
y_{i}}{\hat\theta^{(r)}+1})e^{-\hat\theta^{(r)}
y_{i}}\big)}{\big(1-\hat p^{(r)}(1+\frac{\hat\theta^{(r)}
y_{i}}{\hat\theta^{(r)}+1})e^{-\hat\theta^{(r)} y_{i}}\big)}.
\end{equation*}

\section{Applications to real data sets}
In this Section we fit LG distribution to two real data sets and
compare the fitness with the extended Lindely (Bakoucha et al.,
\cite{Bakoucha2011}), Lindely, Weibull and exponential
distributions, whose densities are given by
\begin{equation*}
\begin{array}{l}
f_{EL}(x;\alpha,\theta,\gamma)=\frac{\theta(1+\theta+\theta
x)^{\alpha-1}}{(1+\theta)^{\alpha}}\big[\gamma(1+\theta+\theta
x)(\theta x)^{\gamma-1}
-\alpha\big]e^{-(\theta x)^{\gamma}},~x,\theta,\gamma>0,~\alpha \in R^{-}\bigcup\{0,1\}, \medskip\\
f_{L}(x;\theta)=\frac{\theta^{2}}{\theta+1} (1 + x) e^{-\theta
x},~x,\theta>0,\medskip\\
f_{WE}(x;\theta,\gamma)=\theta\gamma x^{\gamma-1}e^{-\theta x^{\gamma}},~x,\theta,\gamma>0,\medskip\\
f_{E}(x;\theta)=\theta e^{-\theta x},~x,\theta>0,
\end{array}
\end{equation*}
respectively. The first data set represents the waiting times (in
minutes) before service of 100 bank customers. This data is examined
and analyzed by Ghitany et al. \cite{Ghitany2008} in fitting the
Lindely distribution.

In the second data set, we consider vinyl chloride data obtained
from clean upgradient monitoring wells in mg/L; this data set is
used by Bhaumik et al. \cite{Bhaumik} in fitting the gamma
distribution for small samples.

In order to compare distributions, we consider the K-S
(Kolmogorov-Smirnov) statistic with its respective \textit{p}-value,
$-2\log(L)$, AIC (Akaike Information Criterion), AICC (Akaike
Information Criterion Corrected), BIC (Bayesian Information
Criterion), AD (Anderson-Darling) and CM (Cramer-von Mises)
statistics for the two real data sets.

The best distribution corresponds to lower $-2\log(L)$, AIC, AICC,
BIC, AD and CM statistics values. Table 1 shows parameter MLEs with
the standard errors according to each one of the five fitted
distributions for the two real data sets. Also, Table 2 shows the
values of $-2\log(L)$, K-S statistic with its respective
\textit{p}-value, AIC, AICC, BIC, AD and CM statistics values.

The values in Table 2, indicate that the LG distribution is a strong
competitor to other distributions commonly used in literature for
fitting lifetime data. These conclusions are corroborated by the
fitted pdf, cdf and survival functions of the LG, EL, Lindley,
Weibull and exponential distributions in Fig. 3. We observed a
difference between the fitted curves, which is a strong motivation
for choosing the most suitable distribution for fitting these two
sets of data. From the above results, it is evident that the LG
distribution is the best distribution for fitting these data sets
compared to other distributions considered here.

\begin{table}[tp!]
\caption{MLEs (STDs) of the fitted distributions corresponds to data
1 and 2}
\begin{small}
\begin{tabular}{|c|l|cccc|}
\hline
&Model& $\hat{\alpha}$ & $\hat{\gamma}$  & $\hat{\theta}$ &$\hat{p}$ \\
\hline
&LG & --&--&0.2027(0.0346)&-0.2427(0.5270) \\
&EL & -1e-08(4.3212)&1.4585(0.1098)&0.9128(0.0066)&-- \\
Data 1&Lindely & --&--&0.1866(0.0133)&-- \\
&Weibull & --&1.4585(0.1098)&0.0305(0.0096)&-- \\
&Exp & --&--&0.1012(0.0101)&-- \\
\hline\hline
&LG & --&--&0.5458(0.2305)&0.6346(0.3079) \\
&EL & -1.4435(3.9990)&1.1380(0.4395)&0.2937(0.4138)&-- \\
Data 2&Lindley & --&--&0.8238(0.1054)&-- \\
&Weibull & --&1.0102(0.1327)&0.5202(0.1177)&-- \\
&Exp & --&--&0.5321(0.0913)&-- \\
\hline
\end{tabular}
\end{small}
\end{table}

\begin{table}[h!]
\caption{K-S, \textit{p}-values, $-2\log(L)$, AIC, AICC, BIC, AD and
CM corresponds to data 1 and 2}
\begin{small}
\begin{tabular}{|c|l|cccccccc|}
\hline
&Model& K-S& \textit{p}-value&$-2\log(L)$& AIC  & AICC &BIC& AD& CM \\
\hline
&LG &0.0567 &0.9048&637.8&641.8&642&647.1&0.3984&0.1312 \\
&EL &0.0578 &0.8926&637.5&643.5&643.7&651.3&0.4056&0.1435 \\
Data 1&Lindley &0.0677&0.7486&638.1&640.1&640.1&642.7&0.4865&0.1407 \\
&Weibull &0.0573 &0.8977&637.5&641.5&641.6&646.7&0.4022&0.1425 \\
&Exp &0.1729 &0.0051&658&660&660.1&662.6&4.2237&0.7966 \\
\hline\hline
&LG &0.0800 &0.9814&110.6&114.6&115&117.6&0.2203&0.1108 \\
&EL &0.0813&0.9780&110.6&116.6&117.4&121.2&0.2402&0.1159 \\
Data 2&Lindely &0.1326&0.5880&112.6&114.6&114.7&116.1&0.6873&0.1993 \\
&Weibull &0.0919 &0.9364&110.9&114.9&115.3&118&0.2826&0.1242 \\
&Exp &0.0889 &0.9508&110.9&112.9&113&114.4&0.2719&0.1214 \\
\hline
\end{tabular}
\end{small}
\end{table}
\newpage
\section{Conclusion}
We propose a new two-parameter distribution, referred to as the LG
distribution which contains as special the Lindley distribution. The
hazard function of the LG distribution can be decreasing, increasing
and bathtub-shaped. Several properties of the LG distribution such
as moments, maximum likelihood estimation procedure via an
EM-algorithm, moments of order statistics, residual life function
and probability weighted moments are studied. Finally, we fitted LG
model to two real data sets to show the potential of the new
proposed distribution.


\newpage

\section*{Appendix}
\noindent \textit{Proof of Theorem 1:}\\
The behavior of $f(y)$ is completely similar to the behavior of
$\log(f(y))$. For simplicity we consider the behavior of
$\log(f(y))$. The derivation of $\log(f(y))$ with respect to $y$ is
given by
\begin{equation*}
\frac{\partial}{\partial y}[\log(f(y))]=\frac{e^{\theta y }
(1+\theta)(1-\theta -y \theta )-p(1+(1+y)^2 \theta ^2)}{(1+y)
\left(e^{\theta y}(1+\theta )-p(1+\theta+\theta y)\right)}.
\end{equation*}
The dominator of $\frac{\partial}{\partial y}[\log(f(y))]$ is
positive for each value $y$, $p$ and $\theta$, therefore we only
consider its nominator. Suppose that
\begin{equation*}
g_{1}(y)=e^{\theta y } (1+\theta)(1-\theta -y \theta ),
\end{equation*}
and
\begin{equation*}
g_{2}(y)=p(1+(1+y)^2 \theta ^2).
\end{equation*}
Note that $g_{1}(0)=1-\theta^2$ and $g_{2}(0)=p(1+\theta^2)$.
$g_{2}(y)$ is an increasing function of $y$ for $p\geq0,~\theta>0$
and a decreasing function for all values $p<0,~\theta>0$. Also,
$g_{1}(y)$ is a decreasing function of $y$ for all values $p$ and
$\theta$. Consider the following comparisons between $g_{1}(y)$ and
$g_{2}(y)$:\\
\begin{description}
    \item[(i)] For all values of $p$ and $\theta$ for which $p>
    \frac{1-\theta^2}{1+\theta^2}$, then $g_{1}(y)-g_{2}(y)> 0$ and hence $\frac{\partial}{\partial
    y}[\log(f(y))]<
    0$ which implies the decreasing behavior of $f(y).$
    \item[(ii)] For all values of $p$ and $\theta$ for which
    $p\leq\frac{1-\theta^2}{1+\theta^2}$, then there exist a point $y^{*}>0$ such that (a) for each
    $y<y^{*}$; $g_{1}(y)-g_{2}(y)\leq 0$ and hence $\frac{\partial}{\partial
    y}[\log(f(y))]>0$ which implies the increasing behavior of $f(y)$; (b) for each
    $y\geq y^{*}$; $g_{1}(y)-g_{2}(y)\leq 0$ and hence $\frac{\partial}{\partial
    y}[\log(f(y))]<0$ which implies the decreasing behavior of
    $f(y)$. Thus in this case the pdf $f(y)$ has a unique mode at point
    $y^{*}$ and also is unimodal.
\end{description}

\noindent \textit{Proof of Theorem 2:}\\
The hazard rate function of the LG distribution in (\ref{hazard}) is
given by
\begin{equation*}
h(y)=\frac{\theta^{2}(\theta+1)(1+y)}{(1+\theta+\theta
y)[\theta+1-p(1+\theta+\theta y)e^{-\theta y}]}.
\end{equation*}
The derivation of $\log(h(y))$ with respect to $y$ is given by
\begin{equation*}
\frac{\partial}{\partial y}[\log(h(y))]=\frac{e^{\theta y }
(1+\theta)-p(1+\theta+\theta y)(1+(1+y)^2 \theta
^2)}{(1+y)(1+\theta+\theta y) \left[e^{\theta y}(1+\theta
)-p(1+\theta+\theta y)\right]}.
\end{equation*}
The dominator of $\frac{\partial}{\partial y}[\log(h(y))]$ is
positive for each values $y$, $p$ and $\theta$, therefore we only
consider its nominator. Suppose that
\begin{equation*}
h_{1}(y)=e^{\theta y } (1+\theta),
\end{equation*}
and
\begin{equation*}
h_{2}(y)=p(1+\theta+\theta y)(1+(1+y)^2 \theta ^2).
\end{equation*}
$h_{1}(y)$ is an increasing function of $y$ where
$h_{1}(0)=1+\theta>0$ and $h_{1}(y)>0$ for each $y>0$ and
$\theta>0.$\\
$h_{2}(y)$ is a polynomial function of $y$ which is an increasing
function of $y$ for $0\leq p<1$ and decreasing for $p<0$ and
$h_{2}(0)=p(1+\theta)(1+\theta^2).$ Note that
$$h_{2}^{\prime}(y)=p\theta(3(\theta(1+y))^{2}+2\theta(1+y)+1),$$ and
$$h_{2}^{\prime\prime}(y)=2p\theta^{2}(3(\theta(1+y))+1).$$
Comparison of functions $h_{1}(y)$ and $h_{2}(y)$ implies the
following results:
\begin{description}
    \item[(i)] If $1+\theta<p(1+\theta)(1+\theta^2)$ or
    $p>\frac{1}{1+\theta^2}$, then there exist a unique point $y^{*}>0$ such
    that $h_{1}(y)-h_{2}(y)<0$ for $0<y<y^{*}$ and
    $h_{1}(y)-h_{2}(y)>0$ for $y>y^{*}$. In this case $h(y)$ is
    first decreasing for $0<y<y^{*}$ and then increasing function of
    $y$ for $y^{*}<y$, therefore $h(y)$ is bathtub-shaped.
    \item[(ii)] If $p\leq\frac{1}{1+\theta^2}$, then there exist two points $y^{*}_{1}>0$ and $y^{*}_{2}>0$ such
    that $h_{1}(y)-h_{2}(y)>0$ for $0<y<y^{*}_{1}$,
    $h_{1}(y)-h_{2}(y)<0$ for $y^{*}_{1}<y<y^{*}_{2}$ and $h_{1}(y)-h_{2}(y)>0$ for $y>y^{*}_{2}$. In this case $h(y)$ is
    increasing for $0<y<y^{*}_{1}$, decreasing for $y^{*}_{1}<y<y^{*}_{2}$ and then increasing function of
    $y$ for $y>y^{*}_{2}$, therefore $h(y)$ is firstly increasing
    and then bathtub-shaped, in this case.
\end{description}

\end{document}